\documentclass[%
 reprint,
 amsmath,amssymb,
 aps,
]{revtex4-2}

\usepackage{graphicx}
\usepackage{dcolumn}
\usepackage{bm}
\usepackage{color}
\usepackage[T1]{fontenc}
\usepackage[normalem]{ulem}

\newcommand{\vect}[1]{\boldsymbol{#1}}

\begin{document}

\preprint{APS/123-QED}

\title{Optoelectronic Fingerprints of Interference between Different Charge Carriers\\ in Graphene Superlattices and Analogies to Twisted Graphene Bilayers}%
\author{Sa\'ul A. Herrera}

\author{Gerardo G. Naumis}%
 \email{naumis@fisica.unam.mx}
\affiliation{%
Depto. de Sistemas Complejos, Instituto de F\'isica, \\ Universidad Nacional Aut\'onoma de M\'exico (UNAM)\\
Apdo. Postal 20-364, 01000, CDMX, M\'exico.
}%
\date{\today}

\begin{abstract}
Motivated by recent experimental findings on the low-energy spectrum of Kekul\'e-patterned graphene, the optoelectronic signatures of graphene superlattices with a spatial modulation that triples the size of the unit cell and folds the valleys to the center of the Brillouin zone are studied. For superlattices like those visualized in recent experiments, the optoelectronic response reveals multiple species of carriers distinguished by their effective masses or Fermi velocities. Their signatures are similar to those of systems hosting multifold fermions in which different frequency intervals are dominated by different types of quasiparticles.
Remarkably, the response of these systems exhibits a  characteristic  peak in the optical conductivity suggesting a kind of  interference between the different species of carriers. We also discuss a related superlattice that exhibits merging Dirac cones and band flattening, with a Hamiltonian that resembles a version of the chiral model for twisted bilayer graphene where the long-range moir\'e modulation has been substituted by  a two-parameter bias.
\end{abstract}

\maketitle

\section{Introduction}  
The exotic effects that spatial modulations can induce in the electronic properties of two-dimensional materials
has been the focus of many theoretical and experimental works in the last years \cite{Mudry2007,Sukosin2012,Park2008,Yankowitz2012,Ponomarenko2013,NaumisReview}. More recently, this interest has been further accelerated by the discovery of intriguing correlation phenomena in twisted bilayer graphene (TBG) \cite{Cao2018,Tarnopolsky2019,Morell2010,Bistritzer2011,Ohta2012,Tarnopolsky2019,MOGERA2020470,Wu_2020,Chen2019,Yankowitz2018,Ni2015,Bistritzer,Zheng2016}, where a slight mismatch between two rotated graphene lattices leads to large-scale spatial modulations, known as moir\'e patterns. 
The study of spatial modulations has also played a main role in the field of valleytronics \cite{Schaibley2016,Mordechai2018,Xin2018,Xiao2007,Yao2008,YLee2020}, which focuses in the control of the valley degree of freedom to search for novel mechanisms in quantum transport or for information storage. 

One of the most interesting examples of superlattices resulting from spatial modulation in a 2D material is Kekul\'e-patterned graphene. This phase was first proposed as a novel platform hosting fractionally charged topological excitations \cite{Mudry2007,Chamon2000} and later considered in a mechanism for unconventional superconductivity in graphite \cite{Bitan_2010}. Very recently, Kekul\'e ordering has been predicted to arise in the correlated insulating states of TBG \cite{Bultnick2018,Hoi_2019}, increasing the interest in the study of Kekul\'e-patterned superlattices.

There has also been increasing interest in the transport properties of Kekul\'e-patterned graphene for applications in valleytronics \cite{Wang2015,Ren2015,Wang2018,Elias_2019,Tijerina_2019,Wu2020,Wang2020}, since the symmetry of the modulation folds the $K$, $K'$ valleys to the center of the BZ and enables intervalley transport for low-energy carriers \cite{Cheianov2009,Gamayun,Naumis2020,Herrera2020}.
Kekul\'e ordering has been predicted to arise in graphene due to multiple mechanisms like the ordering of adatoms \cite{Cheianov2009,Gonzalez2018}, substrate mismatch \cite{Giovannetti2015,Gutierrez2016,Wallbank2013}, isotropic strain \cite{Sorella2018}, electron-phonon coupling \cite{Classen} and spin-phonon coupling \cite{Weber2021}.
However, the experimental realization of Kekul\'e-patterned graphene and the probing of its electronic structure was not achieved until very recently \cite{Gutierrez2016,Eom2020,Shuyun2021}. The measurement of the low-energy density of states \cite{Eom2020} supported  predictions about the existence of two Kekul\'e-ordered phases: one preserving the Dirac point and the other opening a gap \cite{Gamayun}. However, further studies are required to support the presence of other important features like the valley-momentum locking \cite{Eom2020,Gamayun}, which refers to a coupling between the momentum $\vect{p}$ and the valley isospin $\tau=K,K'$ introduced by the Kekul\'e order. This coupling is described by an additional term $\vect{p}\cdot\vect{\tau}$ in the Dirac Hamiltonian, analogous to the helicity operator $\vect{p}\cdot\vect{\sigma}$ describing momentum and pseudospin $\sigma=A,B$ coupling in pristine graphene. 

There are three main contributions of this work: (1)
Focusing on the types of superlattices that were recently reported in experiments by Eom et al \cite{Eom2020},
we discuss the optical signatures that might prove useful in their experimental characterization by, for example, confirming the momentum-valley locking \cite{Gamayun}. (2)  We probe the robustness and generality of such signatures by analyzing multiple superlattices. This is important since multiple phases can be present \cite{Eom2020} and because other factors, like second-neighbor interactions \cite{Elias2020} or a substrate-induced ionic potential \cite{Gamayun}, might become important. This also gives information about which signatures are a direct consequence of the symmetry induced by the modulation. (3) We discuss a model for  a related superlattice which, due to the Brillouin zone folding, exhibits merging Dirac cones and presents some qualitative similarities to the process of band flattening in TBG. The mechanism of Brillouin zone folding has been recently demonstrated as an alternative route to TBG for inducing flat bands in a graphene superlattice \cite{Ehlen2020}, and this model might provide an interesting related platform. 

In the following, we introduce the models to be studied and discuss the lattices and their low-energy Hamiltonians, focusing first on superlattices with  the symmetries of those experimentally identified by Eom et al. \cite{Eom2020}. Then, we study the  dynamic polarizabilities and optical conductivities of these systems and find the signatures in their optoelectronic response that might be useful for their experimental characterization, as has been the case for strained graphene \cite{Oliva2016,NaumisReview}. An emphasis is made here on the characteristic signature that could help verify the theoretical prediction of a valley-momentum locking \cite{Gamayun}.
\\~\\
\section{Kekul\'e-patterned graphene} 
\begin{figure}[t]
\hspace*{-.5cm}
\includegraphics[width=0.5\textwidth]{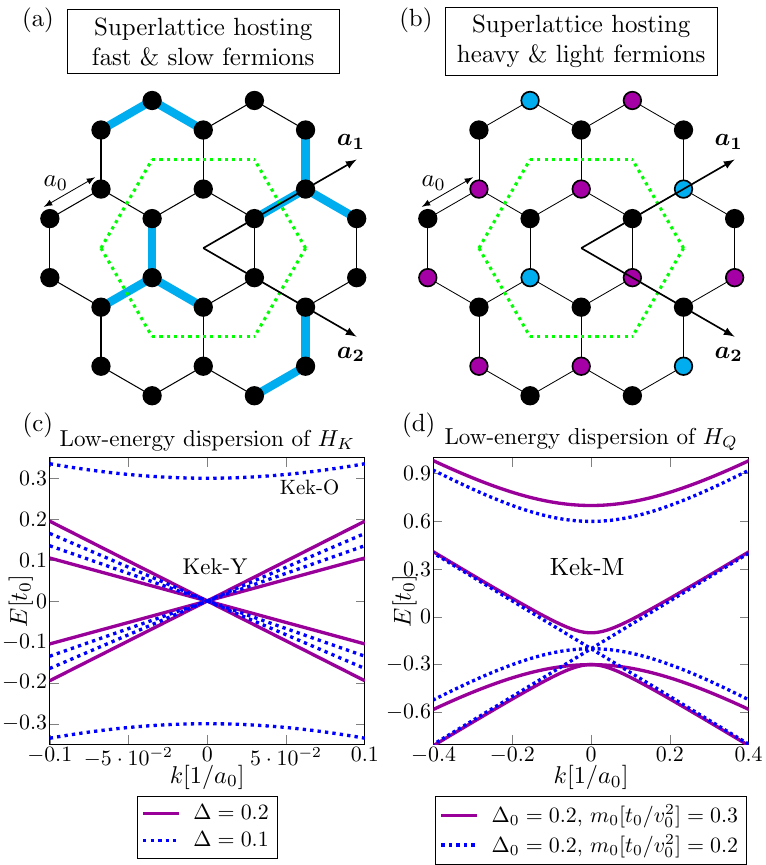}
\caption{\label{Fig:Lattices_and_Dispersions}
Superlattices hosting different species of carriers and their low-energy dispersions. (a) Superlattice associated with $H_K$ (Kek-Y phase). All on-site energies are equal but there are two different bond strengths. The optical response of this superlattice reveals quasiparticles with velocities $v_+$ and $v_-$. (b) Superlattice associated with $H_Q$ (Kek-M phase). All bond strengths are equal but there are three different on-site energies. The optical response of this superlattice reveals quasiparticles with effective masses $m_+$ and $m_-$. In (c) and (d) the  low-energy dispersions of the Kek-Y, Kek-O and Kek-M phases are shown for different values of the parameters.}
\end{figure}

We begin our discussion with the model corresponding to a Kekul\'e-patterned graphene superlattice in which the modulation is introduced by a bond-density wave tripling the size of the unit cell. Fig. \ref{Fig:Lattices_and_Dispersions}a illustrates one of such phases. These bond modulations have been predicted to originate from strain \cite{Sorella2018,Eom2020}, electron-phonon coupling \cite{Classen}, and other mechanisms \cite{Cheianov2009,Weber2021,Giovannetti2015,Gutierrez2016,Wallbank2013,Lin2017}. The low-energy Hamiltonian is given by \cite{Gamayun}, 
\begin{equation}\label{Eq.H_K}
    H_K=
    \begin{pmatrix}
    0 & v_0k_- & \Delta Q_{\nu,+}^*& 0 \\
    v_0k_+ & 0 & 0 & \Delta Q_{\nu,-}^* \\
    \Delta  Q_{\nu,+} & 0 & 0 & v_0k_- \\
    0 & \Delta Q_{\nu,-} & v_0k_+ & 0
    \end{pmatrix},
\end{equation}
acting on the spinor $\Psi=(\psi_{K,A},\psi_{K,B},-\psi_{K',B},\psi_{K',A})$, with $Q_{\nu,\pm}=v_0|\nu|(\nu k_x-ik_y)\pm3t_0(1-|\nu|)$, $k_\pm=k_x\pm ik_y$, $v_0$ is the Fermi velocity in pristine graphene, and the (real) parameter $\Delta$ is the coupling amplitude. The index $\nu=0,\pm1$ leads to the Kek-O phase for $\nu=0$ and the Kek-Y phase for $|\nu|=1$. Recent experiments have supported this model \cite{Eom2020}. The band structures for both Kek-O and Kek-Y exhibit the two valleys folded into the $\Gamma$-point.  As seen in Fig. \ref{Fig:Lattices_and_Dispersions}, the Kek-O phase opens a gap while the Kek-Y phase retains the gapless dispersion \cite{Gamayun}. The band touching in the Kek-Y phase is protected by the threefold rotation symmetry around the sites of one sublattice \cite{Koshino2014}, which is absent in the Kek-O phase. Therefore, the Kek-O phase is not expected to exhibit optical activity for low frequencies and small doping, and thus our discussion will be focused on the Kek-Y phase. Nevertheless, as we discuss below, some results apply to both the Kek-Y and Kek-O phases. The energy dispersion of the Kek-Y phase is
\begin{equation}\label{Eq.E_kek}
E_{k\alpha}^\beta=\alpha (v_0+\beta\Delta v_0)k,
\end{equation}
with $\alpha,\beta=\pm$. Taking $\Delta\rightarrow0$ leads to the case of no modulation (pristine graphene). The low-energy dispersions for the Kek-Y and Kek-O phases are shown in Fig. \ref{Fig:Lattices_and_Dispersions}c. After introducing two sets of Pauli matrices, one for the pseudospin $\sigma_i$ and one for the valley degree of freedom $\tau_i$ ($i=0,x,y,z$), the Hamiltonian for the Kek-Y phase can be written in the compact form $H_K=v_0(\vect{k}\cdot\vect{\sigma})\otimes\tau_0+\Delta v_0\sigma_0\otimes(\vect{k}\cdot\vect{\tau})$, where the second term defines the valley-momentum locking \cite{Gamayun}.
\\~\\
We introduce now a model for a graphene superlattice sharing the same symmetry, and thus also exhibiting a tripled unit cell with the two valleys folded into the $\Gamma$-point. In this model, however, the superlattice is produced due to the on-site energies of the atoms being modulated by, for example, the interaction with a substrate \cite{Venderbos2016,Giovannetti2015} (see Fig. \ref{Fig:Lattices_and_Dispersions}b). The study of this second model will help us to understand how does the optoelectronic response depend on the physical origin of the modulation and to identify the more robust signatures that are inherent to the symmetry.
This model was used in Ref. \cite{Venderbos2016} to study the realization of the quantum anomalous Hall effect in graphene introduced by the influence of a suitable substrate. Also, a similar structure has been predicted for graphene-In$_2$Te$_2$ bilayers \cite{Giovannetti2015}. Moreover, since a substrate-induced ionic potential or second-neighbor interactions (which might become important in experiments) produce a similar patterning in the Kekul\'e phases  \cite{Gamayun,Andrade2021}, these are additional reasons to consider this model. The lattice is shown in Fig. \ref{Fig:Lattices_and_Dispersions}b. It consists of three different onsite energies, with all the bond strengths being the same. The corresponding low-energy Hamiltonian can be written as \cite{Venderbos2016},
\begin{equation}\label{Eq.H_Q}
    H_Q=
    \begin{pmatrix}
    m_0v_0^2 & v_0k_- & 0 & 2 t_0\Delta_0 \\
    v_0k_+ & -m_0v_0^2 & 0 & 0 \\
    0 & 0 &-m_0v_0^2 & v_0k_- \\
    2 t_0\Delta_0 & 0 & v_0k_+ & m_0v_0^2
    \end{pmatrix}
\end{equation}
acting on the same spinor basis, $\Psi=(\psi_{K,A},\psi_{K,B},-\psi_{K',B},\psi_{K',A})$. The parameter $t_0$ is the hopping integral defining the energy scale, $\Delta_0$ introduces a valley coupling and $m_0$, an effective mass, breaks the sublattice symmetry. The low energy dispersion for this Hamiltonian is
\begin{equation}\label{Eq.E_M}
E_{k\alpha}^\beta=\beta\Delta  m_0v_0^2+\alpha\sqrt{v_0^2k^2+(m_0+\beta \Delta m_0)^2 v_0^4},
\end{equation}
with $\alpha,\beta=\pm$ and after scaling the coupling parameter as $\Delta=t_0\Delta_0/m_0v_0^2$ for convenience. Taking $\Delta_0,m_0\rightarrow0$ leads to the case of no modulation (pristine graphene). In the following we refer to this as the Kek-M phase. The corresponding energy dispersion [Eq. (\ref{Eq.E_M})] is shown in Fig. \ref{Fig:Lattices_and_Dispersions}d. For more details on the Hamiltonians see Appendix \ref{App.Hamiltonians}

From the energy dispersions in Eqs. (\ref{Eq.E_kek}) and (\ref{Eq.E_M}) it is easy to see that the index $\alpha$ distinguishes between conduction ($\alpha=+$) and valence ($\alpha=-$) bands, as in the case of pristine graphene ($\Delta\rightarrow 0$).
Due to the valley degeneracy in the $\Delta\rightarrow 0$ case, the description is usually reduced to that of a single-valley Hamiltonian, requiring a single index $\alpha=\pm$ to label the eigenstates. On the other hand, when the Kekul\'e order introduces the valley-coupling $\Delta>0$, the states of both valleys $K,K'$ are considered, and an additional index $\beta$ must to be introduced. Notice however, that the index $\beta$ does not label $K$- and $K'$-polarized states. Instead, it distinguishes between energy dispersions with different Fermi velocities $v_\beta=v_0+\beta\Delta v_0$ (in the the Kek-Y phase) or effective masses $m_\beta=m_0+\beta\Delta m_0$ (in the Kek-M phase). This is already apparent in Eqs. (\ref{Eq.E_kek}) and (\ref{Eq.E_M}). 

In the following section we show that, indeed, the optoelectronic response of these phases (within linear response theory) is that of two species of Dirac quasiparticles with different Fermi velocities $v_\pm$ or effective masses $m_\pm$ plus a term producing an ``interference'' signature, and that such response can be written in terms of single-valley polarizabilities. This shows that, at least in the context of the  optoelectronic response, the Dirac quasiparticle behavior is not completely destroyed by the Kekul\'e order. This is a non-trivial result, since the Dirac quasiparticle picture in graphene is based on a single-valley definition and, in general, a modulation that couples or folds the valleys could destroy such picture.
\\~\\

\section{Optical Conductivity}

The optoelectronic response, within linear response theory, is given by the dynamical polarizability , which can be written as \cite{Peres,Sarma2007,Wunsch}, 
\begin{equation}\label{Eq.Polarizability}
    \Pi(\omega,q)=-g_s\sum_{\alpha\alpha'\beta\beta'}\int\frac{d^2k}{4\pi^2}\frac{f_{k\alpha}^\beta-f_{k'\alpha'}^{\beta'}}{E_{k\alpha}^\beta-E_{k'\alpha'}^{\beta'}+\omega^+}F_{\alpha\alpha'}^{\beta\beta'}(\vect{k},\vect{k'})
\end{equation}
where $f_{k\alpha}^\beta=[\exp(E_{k\alpha}^\beta-\mu)/k_BT+1]^{-1}$ is the Fermi-Dirac distribution, $g_s=2$ is the spin degeneracy and $\omega^+=\omega+i\eta_0$ is the frequency with an infinitesimally small imaginary part added for convergence. The scattering probability is given by the form factor $F_{\alpha\alpha'}^{\beta\beta'}(\vect{k},\vect{k'})=|\langle\Psi_{k\alpha}^\beta|\Psi_{k'\alpha'}^{\beta'}\rangle|^2$ with $\vect{k'}=\vect{k}+\vect{q}$.
\\~\\
\begin{figure}[t]
\hspace*{-.2cm} 
\includegraphics[width=0.49\textwidth]{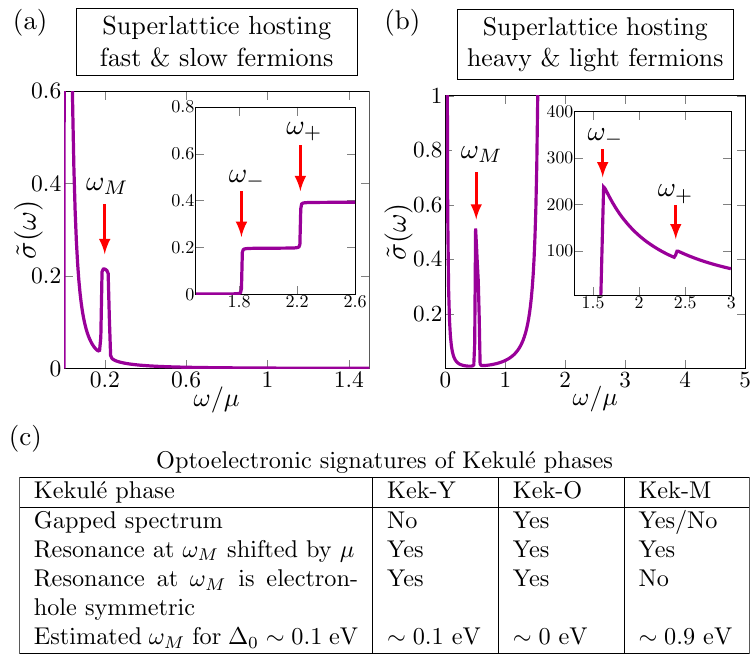}
\caption{\label{Fig:Sigma_abc} Optical conductivity of Kekul\'e superlattices, showing fingerprints of different species of carriers originating from different types of Kekul\'e patterning. (a) Optical conductivity of the Kek-Y phase for a coupling of $\Delta=0.1$. The inset shows different activation frequencies for massless carriers with velocities $v_\pm$  (b) Optical conductivity of the Kek-M phase for the parameters $\Delta_0=0.2$ and $m_0v_0^2/t_0=0.3$. The inset shows different activation frequencies for carriers with effective masses $m_\pm$. The conductivity is shown in units of $4e^2/h$. (c) Summary of the signatures of each phase (estimations of $\omega_M$ correspond to $m_0v_0^2/t_0\sim0.2$ and $\mu\sim0.5$ eV).
}
\end{figure}
In the following we discuss the signatures in the optical conductivity, which can be obtained directly from Eq. (\ref{Eq.Polarizability}) in the limit of $q\rightarrow0$ \cite{Peres,Herrera2020}. In Fig. \ref{Fig:Sigma_abc} we plot the optical conductivities obtained for the Kek-Y and the Kek-M phases using the low-energy models introduced above. Two interesting features are seen to appear in the optical conductivities of both superlattices: (1) Whereas the interband conductivity of pristine graphene starts at an onset frequency of $\omega_0=2\mu$ (due to Pauli blocking), for the two Kekul\'e superlattices two onset frequencies  $\omega_\pm\approx2\mu(1\pm\Delta)$ are seen instead.
(2) An absorption peak arises at low frequencies in the optical conductivity of both superlattices.
The resonance occurs at a frequency given by 
\begin{equation}\label{Eq:Omega_B}
    \omega_M=\frac{\omega_+-\omega_-}{2}.
\end{equation}
Interestingly, this last relation coincides with the expression for the frequency of a pattern arising from the interference of two slightly-mismatching spatial or temporal scales defined by frequencies $\omega_+$ and $\omega_-$. In fact, the periodicity of the large-scale moir\'e patterns that arise in moir\'e superlattices are given by analogous expressions. Because of this, we refer to the resonance at $\omega_M$ as an ``inteference'' signature. We make the remark that the relation in Eq. (\ref{Eq:Omega_B}) holds for both models regardless of the fact that $H_K$ and $H_Q$ describe modulations with different physical origins, have different energy dispersions, and that the expressions for $\omega_\pm$ and $\omega_M$ as a function of the valley coupling are different in each case. This points to the signature being originated from the symmetry alone. In terms of coupling parameters, the resonance peak for the Kek-M phase is given by $\omega_M\approx2\Delta_0t_0$ (at high doping). For the case of the Kek-Y phase, the peak occurs at $\omega_M\approx2\Delta\mu$. The resonance at $\omega_M$ corresponds to optical transitions between the upper bands (assuming $\mu>0$), which in pristine graphene correspond to different valleys. Since in pristine graphene these transitions are completely absent, its observation would provide evidence for the predicted valley-momentum locking \cite{Gamayun,Eom2020}.

Remarkably, the features at $\omega_\pm$ in the optical conductivity belong to the response of two species of quasiparticles in each superlattice: massless Dirac fermions with Fermi velocities $v_\pm=v_0\pm\Delta v_0$ in the Kek-Y phase and Dirac fermions with masses $m_\pm=m_0\pm\Delta m_0$ for the Kek-M phase. We refer to the Dirac quasiparticles with velocities $v_+$ and $v_-$ in the Kek-Y phase as ``fast'' and ``slow'' fermions and to the Dirac quasiparticles with masses $m_+$ and $m_-$ in the Kek-M as ``heavy'' and ``light'' fermions, respectively.

To better illustrate that the optoelectronic response corresponds to two species of Dirac quasiparticles in each Kekul\'e phase, we show that the full polarizability can be expressed in terms of the same response functions that correspond to Dirac fermions in pristine graphene. Specifically, the polarizability is given by the sum of the responses of two species of Dirac fermions plus a term describing transitions between their energy dispersions, which produces the interference signature at $\omega_M$. In order to see this, we use the fact that  the full scattering probability $F_{\alpha\alpha}^{\beta\beta'}$ for Kekul\'e-patterned graphene can be written in terms of the single-valley scattering probability $F_{\alpha\alpha'}$ used in the calculation \cite{Wunsch,Sarma2007} of the (single-valley) polarizability of pristine graphene (for details see Appendix \ref{App.P_separation}): 
\begin{equation}\label{Eq:Fq_H1}
   F_{\alpha\alpha'}^{\beta\beta'}(\vect{k},\vect{q})=\delta_{\beta,\beta'}F_{\alpha\alpha'}(\vect{k},\vect{q})-\beta\beta'\Big(\frac{q\sin\varphi}{2|\vect{k+q}|}\Big)^2.
\end{equation}
For the superlattices introduced above, this property allows us to separate $\Pi(\omega,q)$ into three contributions when summing over the $\beta,\beta'$ indices. 

For pristine graphene, the valleys are degenerated and separated in momentum space. Therefore, the total polarizability for low-energy carriers in graphene is simply given by two times (accounting for valley degeneracy) the single-valley polarizability, $\Pi^g_{v_0}(\omega,q)$ \cite{Sarma2007,Wunsch,Peres}. That is,
\begin{equation}\label{Eq.Polarizability_g}
    \Pi(\omega,q)=2\times\Pi^g_{v_0}(\omega,q)\text{\qquad(graphene),}
\end{equation}
where the subscript stands for a Fermi velocity $v_0$ in the energy dispersion  $E=v_0k$ of graphene. Equivalently, $\Pi_{v_0}^g(\omega,q)$ can be understood as the polarizability for massless Dirac fermions with Fermi velocity $v_0$. When a spatial modulation that couples the valleys is introduced Eq. (\ref{Eq.Polarizability_g}) no longer holds, since new terms accounting for electronic transitions between bands that corresponded to different valleys are now possible.  Furthermore, the coupling could destroy the Dirac quasiparticle picture and then the polarizability would not be given solely, or even partially, by $\Pi_{v_0}^g(\omega,q)$. It can be shown, however, by using Eqs. (\ref{Eq.Polarizability}) and (\ref{Eq:Fq_H1}) (see Appendix \ref{App.P_separation}) that the total polarizability of the Kek-Y phase, $\Pi_Y(\omega,q)$, can be written as
\begin{equation}\label{Eq.Polarizability_kek}
    \Pi_Y(\omega,q)=\Pi^g_{v_+}(\omega,q)+\Pi^g_{v_-}(\omega,q)+\Pi^M_{v_M}(\omega,q), 
\end{equation}
where the first two terms on the right side correspond to the same polarizabilities for massless Dirac fermions $\Pi_{v_0}^g(\omega,q)$, only with the original Fermi velocity $v_0$ replaced by a different velocity $v_\pm=v_0\pm\Delta v_0$ in each term, indicating thus that the Kekul\'e order not only preserves the Dirac quasiparticle picture but also leads to two species of carriers with different Fermi velocities. On the other hand, the last term accounts for transitions between the upper bands (which in pristine graphene correspond to bands in different valleys, and therefore such transitions are forbidden)  and is responsible for the interference signature at $\omega_M$  in the optical conductivity (Fig. \ref{Fig:Sigma_abc}a), while the terms $\Pi_{v_\pm}^g$ produce the features at $\omega_\pm$, which are the activation frequencies for the quasiparticles with Fermi velocities $v_\pm$ (see Appendix \ref{App.Sigma}).

For the Kek-M phase, although the physical origin of the modulation and the energy spectrum are different, a completely analogous result is obtained. We find that the total polarizability can be written as
\begin{equation}\label{Eq.Polarizability_M}
    \Pi_Q(\omega,q)=\Pi^g_{m_+}(\omega,q)+\Pi^g_{m_-}(\omega,q)+\Pi^M_{m_M}(\omega,q). 
\end{equation}
where the first two terms on the right side of the last equation correspond to the single-valley polarizabilities for massive (rather than massless) Dirac fermions with an effective mass $m_0$, $\Pi_{m_0}^g(\omega,q)$, only with the original effective mass $m_0$ replaced by a different mass $m_\pm=m_0\pm\Delta m_0$ in each term (one has to consider $\Pi_{m_0}^g(\omega,q)$ instead of $\Pi_{v_0}^g(\omega,q)$ when a gap is induced in the dispersion of graphene by a broken sublattice symmetry \cite{Pyatkovskiy2008}). In this case too, the last term accounts for transitions between the upper bands and is responsible for the resonance at $\omega_M$ in the optical conductivity (Fig.  \ref{Fig:Sigma_abc}b), while the terms $\Pi_{m_\pm}^g$ produce the features at $\omega_\pm$, which can be interpreted as the activation frequencies for the quasiparticles with effective masses $m_\pm$ (see Appendix \ref{App.Sigma}).

Therefore, even though the Kekul\'e order couples and folds the valleys through different types of spatial modulations in the Kek-Y and Kek-M phases, in both cases the full polarizability can be separated into
the response of two species of Dirac quasiparticles plus an additional term that describes the electronic transitions between their energy dispersions and produces an interference signature.

Although the interference signature at $\omega_M=\frac{1}{2}(\omega_+-\omega_-)$ is determined by the activation frequencies $\omega_\pm$ for the two species of quasiparticles, it should be noted that this signature does not arise from the interference of the simultaneous responses of each specie of quasiparticle. Consider, for example, probing the material with an incident field of frequency $\omega=\omega_M$. Because $\omega_M<\omega_\pm$, this would produce the resonance even when the response of each  quasiparticle (which occurs at the higher frequencies $\omega_\pm$) is absent.

In summary, the valley coupling introduced by the Kekul\'e order preserves the Dirac quasiparticle picture, while also introducing a splitting of a dynamical property $\mu_0$ (here, it can be the Fermi velocity $v_0$ or the effective mass $m_0$) which splits as $\mu_0\rightarrow \mu_\pm=\mu_0\pm\Delta\mu_0$ when the valley coupling $\Delta$ is introduced. This leads to the total polarizability being given by the sum of the polarizabilities for two species of carriers $\Pi^g_{\mu_\pm}$ plus an additional term $\Pi^M_{\mu_M}$ as,
\begin{equation}
    2\times\Pi^g_{\mu_0}\xrightarrow{\text{$\Delta>0$}}  \Pi^g_{\mu_+}+\Pi^g_{\mu_-} +\Pi^M_{\mu_M}, \nonumber
\end{equation}
where the last term introduces an interference signature at a frequency $\omega_M$, which is determined by the activation frequencies of the new species of quasiparticles. 
\\~\\
\section{Multifold fermions} Some of the signatures discussed so far, namely, different coexisting quasiparticles characterized by different activation frequencies, low-frequency sharp absorption peaks and (in the case of the Kek-Y phase) a multi-step conductivity with a dependence $\sigma\sim\omega^{d-2}$ (where $d$ is the spatial dimension),  are quite similar to those found in the optical conductivity of multifold fermions \cite{Bradlyn2016,Chang2017}.
Multifold fermions are the generalization of Weyl fermions to a higher effective spin representation that exhibit a remarkable optoelectronic response, including exotic circular photogalvanic effects \cite{Chang2017,deJuan2017,Ma2017,Takane2019,Ni2021}. A number of crystals have recently been shown to exhibit multiple species of these quasiparticles coexisting at low energies and, particularly, the study of their optical conductivity  has been the focus of multiple theoretical and experimental works \cite{Takane2019,Maulana2020,Chang2017,Flicker2018,Martinez2019,Xu2020,Tetsuro2019,Ni2020,Li2019}. 
It has been noted in previous works \cite{Giovannetti2015,Venderbos2016,Herrera2020} that, due to the folding of the $K$ and $K'$ valleys into the $\Gamma$-point, the resulting low-energy band structure in some Kekul\'e-modulated superlattices can be described by higher pseudospin representations of the Dirac equation \cite{Dora2011}. As can be seen in Fig. \ref{Fig:Lattices_and_Dispersions}, the dispersion of the Kek-Y is very similar to that of a pseudospin-$3/2$ system \cite{Herrera2020}, while the dispersion of Kek-M phase resembles that of a  pseudospin-$1$ system \cite{Giovannetti2015,Venderbos2016} (notice the threefold crossing shown in Fig. \ref{Fig:Lattices_and_Dispersions}d). 
Therefore, it could be expected that the optical signatures of the superlattices studied here would share some similarities with those found in systems hosting multifold fermions. Indeed, in systems hosting multifold fermions the optical conductivity exhibits multiple linear steps (characteristic of linearly-dispersive bands $\sigma\sim \omega^{d-2}$), with different activation frequencies for each type of multifold fermion \cite{Martinez2019,Ni2020,Xu2020}. Here, similarly, we find for both the Kek-Y and Kek-M phase different species of carriers exhibiting distinct activation frequencies (features at $\omega_\pm$ in Fig. \ref{Fig:Sigma_abc}a,b). Furthermore,
the optical conductivity of materials hosting multifold fermions like CoSi \cite{Xu2020}, RhSi \cite{Ni2020} and other Weyl semimetals like NbP \cite{Neubauer2018} exhibit low-frequency  narrow peaks originating from transitions between SOC-split bands and the position of such peaks is a measure of the SOC strength \cite{Martinez2019,Xu2020}. Also, the SOC  is responsible for introducing multiple species of quasiparticles  (e.g. by splitting a threefold node into a spin-$3/2$ fermion and a twofold Weyl fermion \cite{Xu2020}). Similarly, we find that very similar sharp peaks appear (around $\omega_M\sim\Delta$) in the conductivity of the Kek-Y and Kek-M phases at low frequencies due to transitions between bands that are split by the valley coupling $\Delta$ introduced by the Kekul\'e modulation, which also introduces the different species of quasiparticles. In both cases the frequency of the sharp peak is given by the coupling amplitude. This suggests that the Kekul\'e modulation in these systems might play a role in the optical response similar to that played by the SOC in systems hosting multifold fermions. These remarks might lead to interesting connections to multifold fermions and deserve further study.
\\~\\
\begin{figure}[t]
\hspace*{-.5cm}
\includegraphics[width=0.5\textwidth]{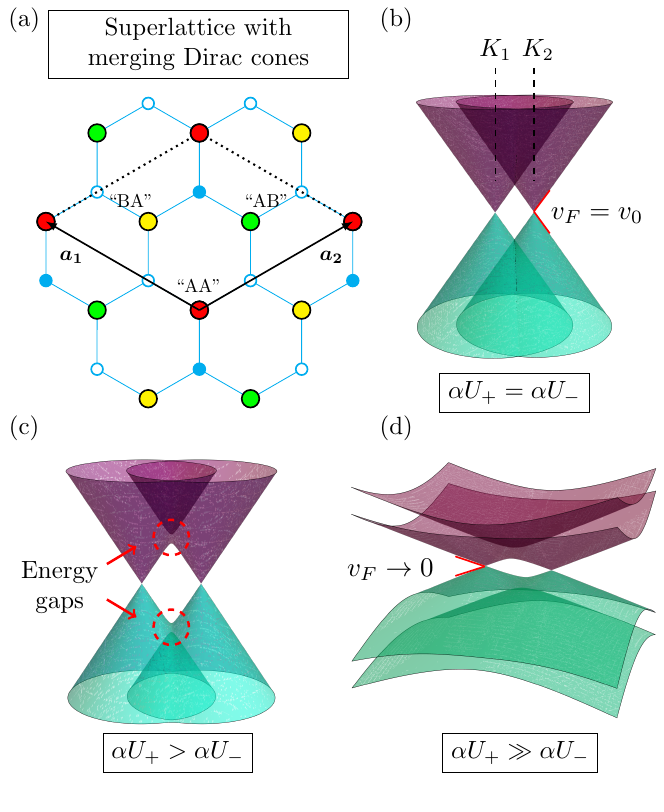}
\caption{\label{Fig:Superlattice_flat_bands}
Merging of the Dirac cones in a $\sqrt{3}\times\sqrt{3}$ graphene superlattice, reminiscent of the qualitative description for intervalley hybridization in magic angle TBG. (a) Graphene superlattice exhibiting merging Dirac cones. The atoms have been labeled in analogy to the special points in TBG, {\color{black} with respective onsite energies $V_{AA}$ and $V_{AB/BA}$ (see text)}. (b) Low-energy spectrum of $\mathcal{H}$ exhibits two Dirac cones at $K_1$, $K_2$. (c) As $U_+$ is increased over $U_-$ the cones start to hybridize. (d) As $U_+$ is further increased (leading to localization in the ``AA'' sites), cone hybridization further flattens the bands, decreasing the Fermi velocity.}
\end{figure}

\section{Kekul\'e Superlattice with Merging Dirac Cones}
The discoveries on twisted bilayer graphene (TBG) have greatly motivated the study of the rich physics related to weakly-dispersive or ``flat'' bands \cite{Tarnopolsky2019,Ledwith2020,Wang2021,Naumis_Chiral2021} and, more recently, {\color{black}there has been an ongoing search for alternative routes to
induce a phenomenology analogous to that of TBG in spatially-modulated single graphene sheets \cite{Skurativska2021,Mao2020,Ehlen2020} and other types of honeycomb structures \cite{Lee2020,Gardenier2020}.
In order to induce flat bands, some proposals have focused on engineering the graphene superlattices by buckling \cite{Mao2020,Milovanovi2020}, or by introducing
a tailored periodic potential \cite{Skurativska2021,Ehlen2020,Naumis_Taboada_2014} which leads to a momentum-space description in a reduced Brillouin zone, also called Brillouin zone folding.}

In this section we discuss a Kekul\'e superlattice in which the Brillouin zone folding leads to the electronic dispersion exhibiting two close Dirac cones that hybridize or ``merge'' {\color{black}as the onsite potential is tuned to induce localization in a triangular sublattice}. We show that the Hamiltonian for this model resembles a version of the chiral model for TBG where {\color{black}the long-range moir\'e modulation has been substituted by a two-parameter Kekul\'e coupling}, and also highlight some {\color{black}qualitative similarities to the band evolution in TBG} that occurs as interlayer tunneling is turned on at a magic angle. {\color{black} In the following, we introduce the model and then discuss its relevance in the context of recent related works, mainly Refs. \cite{Skurativska2021,Ehlen2020,Shuyun2021}.}

We focus on a more general form of the Hamiltonian previously introduced in Eq. (\ref{Eq.H_Q}). As discussed before, it describes a graphene superlattice where a periodic potential triples the size of the unit cell ($a_0\times a_0\rightarrow \sqrt{3}a_0\times\sqrt{3}a_0$) by altering the on-site atomic energies, leading to a unit cell of six (rather than two) carbon atoms (for more details see Appendix \ref{App.Hamiltonians}). The general Hamiltonian has the following form,
\begin{equation}\label{Eq.H_Q_general}
    H_Q=
    \begin{pmatrix}
    m_0v_0^2 & v_0k_- & 0 & t_0\Delta_A \\
    v_0k_+ & -m_0v_0^2 & -t_0\Delta_B^* & 0 \\
    0 & -t_0\Delta_B & -m_0v_0^2 & v_0k_- \\
    t_0\Delta_A^* & 0 & v_0k_+ & m_0v_0^2 
    \end{pmatrix},
\end{equation}

acting in the same basis as in Eq. (\ref{Eq.H_Q}), which is a particular case of this Hamiltonian. We take $m_0\rightarrow 0$ and {\color{black} rewrite $t_0\Delta_A^*=\alpha U_-$ and $t_0\Delta_B=-\alpha U_+$. Therefore, $\alpha=t_0$ defines the energy scale and $U_\pm$ is a two-parameter field (given in terms of the onsite energies of the lattice) that couples the Dirac cones}. After re-shuffling the third and fourth elements of the basis, one gets
\begin{equation}\label{Eq.H_D}
 \mathcal{H}=
    \begin{pmatrix}
    0 & \mathcal{D}_-^* \\
    \mathcal{D}_+ & 0
    \end{pmatrix},
\quad
\mathcal{D}_r=
    \begin{pmatrix}
    -2i\overline{\partial} & \alpha U_r \\
    \alpha U_{-r} & -2i\overline{\partial}
    \end{pmatrix},
\end{equation}
where $r=\pm$ and we have used $k_j\rightarrow -i\partial_{r_j}$ so     $k_+\rightarrow -i(\partial_x+i\partial_y)\equiv -2i\overline{\partial}$. This Hamiltonian resembles a version of chiral model for TBG \cite{Tarnopolsky2019}
where the field $U(\pm r)$ (which couples the top and bottom layers) has been replaced by {\color{black}two coupling amplitudes $U_\pm$, which are determined by the onsite energies of the lattice. In Fig. \ref{Fig:Superlattice_flat_bands}a we show the $\sqrt{3}\times\sqrt{3}$ graphene superlattice with the atomic sites labeled in correspondence to the special points AA, AB/BA in TBG to highlight this analogy.}

{\color{black}Recently, there has been a number of proposals for describing TBG by emergent honeycomb lattices  \cite{Venderbos2018,Yuan2018,Koshino2018,Vafek2018} lacking the long-range modulations but retaining the appropriate symmetries. However,} the Hamiltonian in Eq. (\ref{Eq.H_D}) describes a system that is quite different to (and much simpler than) TBG. {\color{black}The main differences rely not only on the removal of the long-range spatial dependence of the field $U(\vect{r})$ that couples the layers}, but also in the absence of crucial symmetries \cite{Ledwith2021}  {\color{black}(e.g., in TBG the coupling occurs between Dirac cones with the same chirality, while in the present model they possess opposite chirality).  Therefore, we do not consider this system as a model for TBG. Despite this, within the context of engineered graphene monolayers exhibiting a phenomenology analogous to that of TBG \cite{Mao2020,Ehlen2020,Skurativska2021}, it is interesting to consider the evolution of the band dispersion of $\mathcal{H}$ as $U_+$ and $U_-$ are varied. We take a look at the evolution of the band structure when tuning the values of the onsite energies $V_{AA}$ and $V_{AB/BA}$ in such a way that the localization in the lattice mimics the wavefunction of TBG at the first magic angle. In such condition, the wavefunction heavily localizes in the AA sites and presents nodes on the AB/BA sites, as AA stacking disfavors tunneling between layers 
\cite{Tarnopolsky2019}. We therefore take $V_{AA}\rightarrow-\infty$ and $V_{AB/BA}\rightarrow +\infty$. Since the parameters $U_\pm$ are defined in terms of the onsite energies, this choice leads to the condition $U_+\rightarrow\infty$. Notice that such limit is not as trivial as simply favoring the localization in the AA sites. Such limit is not possible because the condition $m_0\rightarrow0$ imposed in Eq. (\ref{Eq.H_D}) requires the localization in the AB/BA sites to be proportionally disfavored (see Appendix A). As $U_+$ increases over $U_-$ (we assume $U_-$ to be constant), the localization in the AA sites leads to the hybridization of the Dirac cones.} The dispersion is shown in Fig. \ref{Fig:Superlattice_flat_bands}b-d.   When $U_+\gg U_-$ the Fermi velocity approaches zero as $v_F\approx2\sqrt{U_-/U_+}$. {\color{black}Importantly, in addition to the flattened bands concentrating its spectral weight around the Fermi level, they are also separated from the other bands in the spectrum.}
This band evolution is reminiscent of the qualitative description that has been used  to describe the process of band flattening and localization in TBG at a magic angle \cite{Cao2018-2}. Beginning with two uncoupled rotated layers, the low-energy band structure consists of the Dirac cones from each layer rotated about the center of the Brillouin zone and forming pairs. As the layers get closer and become coupled, the pairs of cones start to hybridize. The first experimentally verified consequences of this process were the opening of energy gaps at the intersection of the Dirac cones, and a renormalization of the Fermi velocity \cite{Brihuega2012,Li2010,Luican2011,Cao2018-2}, which lead to the flattening of the bands and to localization in a triangular superlattice formed by the moir\'e pattern.

{\color{black} The ongoing search for systems with electronic properties similar to those of TBG, has recently lead to novel proposals based on single graphene sheets with engineered spatial modulations \cite{Skurativska2021,Ehlen2020}.
We highlight Ref. \cite{Skurativska2021}, where flat bands with nontrivial topology where shown to arise in the dispersion of single graphene sheets with a periodic potential induced by adatoms. The system studied therein is highly related to the model discussed in this section and in fact, a similar technique involving the periodic arrangement of adatoms was recently employed to induce  Kekul\'e ordering in graphene \cite{Shuyun2021}. Crucially, the periodic potential proposed in Ref. \cite{Skurativska2021} folds the $K$ and $K'$ points of graphene to the $\Gamma$-point like a Kekul\'e modulation. Such folding allows for the hybridization of the graphene with the adatom bands, leading to the flat bands. The periodic potential leading to such configuration is given by the lattice vector $\vect{v}_1=n\vect{u}_1+(3m+n)\vect{u}_2$ and its 60-degree rotation, where $n,m\in\mathbb{Z}$ and $\vect{u}_{1,2}$ are the lattice vectors of graphene.
The superlattice studied in Ref. \cite{Skurativska2021} corresponds to the case with $(n,m)=(-1,2)$, which leads to a supercell with 42 atoms. We point out that a Kekul\'e superlattice corresponds to the case with $(n,m)=(1,-1)$, which leads to the smallest supercell configuration for such a potential. We also point out that experimental evidence was recently reported \cite{Ehlen2020} for the formation of flat bands in a related system consisting of a graphene sheet with a $2\times2$ superlattice potential (analogous to the $\sqrt{3}\times\sqrt{3}$ potential in Kekul\'e-modulated graphene) induced by layers of cesium atoms. \\

Because Kekul\'e-modulated graphene belongs to the class of superlattices predicted to exhibit topologically nontrivial flat bands in Ref. \cite{Skurativska2021}, and because its synthesis via a periodic arrangement of adatoms was recently demonstrated \cite{Shuyun2021}, it might provide a potential platform to explore electronic behavior analogous to that of TBG in periodically-modulated graphene monolayers. Although the cone hybridization and band flattening in the model discussed in this section are induced solely via the tuning of the onsite energies of the lattice, a more sophisticated version of the model considering the hybridization of the graphene with the adatom bands might potentially lead to topologically nontrivial flat bands like those found in Ref. \cite{Skurativska2021}.
We hope that the discussion in this section further motivates its exploration.}
\\~\\
\section{Conclusion}

We studied graphene superlattices with a tripled unit cell and folded Dirac cones, some of which were visualized in recent experiments. We used linear response theory to find signatures that could aid in the experimental confirmation of recent theoretical predictions as, for example, the valley-momentum locking. We analyzed the robustness of such signatures and their origin. The optical response suggests two species of carriers with signatures similar to those of multifold fermions.
{\color{black}Finally, we introduced a model for a Kekul\'e superlattice that exhibits a dispersion with hybridizing Dirac cones and discussed some of its features in the context of recent proposals for periodically-modulated graphene monolayers exhibiting a phenomenology similar to that of twisted graphene bilayers.} Since two of the graphene superlattices we studied have been recently visualized in experiments (the Kek-Y and Kek-O phases)  \cite{Eom2020,Shuyun2021}, we hope that some of the signatures discussed here can serve to further validate the predicted electronic properties of these systems. 

\begin{acknowledgments}
We thank UNAM-DGAPA project IN102620 and CONACyT project 1564464. S. A. H. was supported by the Consejo Nacional de Ciencia y Tecnolog\'ia (CONACyT). 
\end{acknowledgments}

\appendix
\section{Low-energy models}\label{App.Hamiltonians}
In this appendix we describe with more detail how the Hamiltonians discussed in the main text have been obtained. Two different Hamiltonians were used in this work. Both are based on low-energy approximations of tight-binding models describing a graphene superlattice with a modulation that triples the size of the unit cell ($a_0\times a_0\rightarrow \sqrt{3}a_0\times\sqrt{3}a_0$), also generically called a Kekul\'e distortion \cite{Wallbank2013}, leading to a cell of six carbon atoms.

The Hamiltonian $H_K$ in Eq. (\ref{Eq.H_K}) describing the Kek-O and Kek-Y phases (for $\nu=0$ and $|\nu|=1$, respectively) was derived in Ref. \cite{Gamayun}. It is based on the otherwise usual tight binding model for a graphene lattice,
\begin{equation}
    H=-\sum_{\vect{r}}\sum_{l=1}^3 t_{\vect{r},l}\hat{a}_{\vect{r}}^\dagger \hat{b}_{\vect{r}+\vect{s}_l}+\text{H.c.},
\end{equation}
with the exception that the nearest neighbor (NN) hopping amplitude $t_{\vect{r},l}$ describes the bond-density wave that forms the Kek-Y or Kek-O textures. The vectors $\vect{s}_l$ are the usual vectors connecting the NNs with bond-lengths $a_0$, and the fermionic operator $\hat{a}_{\vect{r}}$ ($\hat{b}_{\vect{r}}$) annihilates an electron at position $\vect{r}$ in the $A$ ($B$) sublattice of graphene. The hopping amplitude is given by
\begin{equation}
    t_{\vect{r},l}/t_0=1+\Re[\Delta e^{i(p\vect{K}_++q\vect{K}_-)\cdot\vect{s}_l+i\vect{G}\cdot\vect{r}}],
\end{equation}
where $t_0$ is the hopping amplitude of pristine graphene and the Kekul\'e wave vector $\vect{G}=\vect{K}_+-\vect{K}_-$ couples the Dirac points at $\vect{K}_\pm$. The velocity $v_0=3|t_0|a_0/2$ in the main text is defined as usual (with $\hbar\equiv1$) and the coupling parameter $\Delta$ has been chosen to be real ($\Delta\rightarrow 0$ leads to the model for pristine graphene). The parameter $\nu=1+q-p \mod 3$ distinguishes between the Kek-Y and Kek-O phases. The low-energy Hamiltonian is obtained after linearizing near  $\vect{k}=0$ and projecting out two high-energy bands leading to a $4\times4$ Hamiltonian. The basis used in Ref. \cite{Gamayun} is $\Psi=(-\psi_{K',B},\psi_{K',A},\psi_{K,A},\psi_{K,B})^T$, known as the valley-isotropic representation. Here, we have interchanged the order of the valleys to keep consistency with the other models, leading to $\Psi=(\psi_{K,A},\psi_{K,B},-\psi_{K',B},\psi_{K',A})^T$.

The Hamiltonian $H_Q$, used to describe the Kek-M phase in Eq. (\ref{Eq.H_Q}) and the superlattice with merging Dirac cones in Eq. (\ref{Eq.H_Q_general}), was derived in Ref. \cite{Venderbos2016}. It consists of a tight binding model for a graphene superlattice where a substrate-induced potential triples the size of the unit cell by altering the on-site atomic energies, leading to a unit cell of six carbon atoms labeled by $A_\alpha$, $B_\alpha$, with $\alpha=1,2,3$. The tight binding model is,
\begin{multline}
    H=-\sum_{\langle\alpha i,\beta j\rangle}t_0 \hat{a}_{\alpha i}^\dagger \hat{b}_{\beta j } + \text{H.c.}\\
    +\sum_{\alpha=1}^3\sum_i(V_{A_\alpha} \hat{n}_{\alpha i}^A+V_{B_\alpha} \hat{n}_{\alpha i}^B),
\end{multline}
where in the first term $t_0$ is the NN hopping amplitude, the fermionic operator $\hat{a}_{\alpha i}$ ($\hat{b}_{\alpha i}$) annihilates an electron at the cell $i$ in the sublattice $A_\alpha$ ($B_\alpha)$, and $\langle\alpha i,\beta j\rangle$ denotes the sum over all the NN. In the second term the $V_{A_\alpha}$ are the onsite energies and $\hat{n}_{\alpha i}^A=\hat{a}_{\alpha i}^\dagger\hat{a}_{\alpha i}$, with the same for $B_\alpha$.  The onsite energies are modeled by a superlattice potential with triangular symmetry, $V(\vect{r})=\sum_{\vect{G}}V_{\vect{G}}e^{i\vect{G}\cdot\vect{r}}$. A first set of vectors introduce a triangular lattice $\vect{G}/G=\{\pm1,0\},\{\pm\cos\frac{\pi}{3},\pm \sin\frac{\pi}{3}\}$ ($G=4\pi/3\sqrt{3}a_0$) of three times the size of the unit cell, while a second set $\vect{\tilde{G}}/\sqrt{3}G=\{0,\pm1\},\{\pm\cos\frac{\pi}{6},\pm\sin\frac{\pi}{6}\}$ breaks the sublattice symmetry. In the main text we have used the same basis as for the $H_K$ Hamiltonian, $\Psi=(\psi_{K,A},\psi_{K,B},-\psi_{K',B},\psi_{K',A})^T$, following Ref. \cite{Beenakker2018}. In this basis, the parameters of Eq. (\ref{Eq.H_Q_general}) as a function of the on-site energies are given as $6m_0v_0^2=\sum_n(V_{An}-V_{Bn})$, $6t_0\Delta_A=2V_{A_1}-V_{A_2}-V_{A_3}+i\sqrt{3}(V_{A_2}-V_{A_3})$, $6t_0\Delta_B=2V_{B_1}-V_{B_2}-V_{B_3}+i\sqrt{3}(V_{B_2}-V_{B_3})$ \cite{Beenakker2018}. There is an additional shift in the diagonal terms given by $V_0=\sum_n(V_{An}+V_{Bn})/6$, but the zero of energy can always be shifted such that $V_0=0$. Eq. (\ref{Eq.H_Q}) is the particular case with $\Delta_B=0$ and $\Delta_A\equiv2\Delta_0$. {\color{black}In Fig. \ref{Fig:Superlattice_flat_bands}a and the discussion after Eq. (\ref{Eq.H_D}) the onsite energies $V_{B_1}$, $V_{B_2}$ and $V_{B_3}$ have been referred to as $V_{AA}$, $V_{AB}$ and $V_{BA}$ in analogy to the special points in TBG and the $U_\pm$ have been assumed real (by taking $V_{A_2}=V_{A_3}$, $V_{B_2}=V_{B_3}$) for simplicity. After this, one has $U_+=[-2V_{AA}+V_{AB}+V_{BA}-i\sqrt{3}(V_{AB}-V_{BA})]/6\alpha$. Notice that the condition $m_0\rightarrow0$ restricts the values of the onsite energies and therefore $V_{AA}$, $V_{AB/BA}$ can not be chosen arbitrarily.
Although we have considered the simplest case of coupling amplitudes $U_\pm$ without a spatial dependence, in general one might define coupling amplitudes $U_\pm(r)$ that vary slowly in space. Such field could be chosen to have the same spatial dependence as the interlayer coupling field $U(r)$ in TBG. An alternative approach to introduce a spatial dependence is to consider a piece-wise coupling $U_\pm$ \cite{Beenakker2018}.  We leave such exploration for further work.}

\section{Separation of the polarizability}\label{App.P_separation}
In this appendix we show how to arrive at Eq. (\ref{Eq:Fq_H1}) and the expression for $\Pi_Y(\omega,q)$ in Eq. (\ref{Eq.Polarizability_kek}). The expression for $\Pi_Q(\omega,q)$ is obtained in a completely analogous way.

We begin with the single-valley polarizability of pristine graphene, $\Pi_{v_0}^g(\omega,q)$. Since the valleys in pristine graphene are decoupled, its total polarizability is given by two times (accounting for valley degeneracy) the single valley-polarizability [Eq. (\ref{Eq.Polarizability_g})], which is then given by,
\begin{equation}\label{Eq.App.P_single_valley}
    \Pi_{v_0}^g(\omega,q)=-g_s\sum_{\alpha\alpha'}\int\frac{d^2k}{4\pi^2}\frac{f_{k\alpha}-f_{k'\alpha'}}{E_{k\alpha}-E_{k'\alpha'}+\omega^+}F_{\alpha\alpha'}(\vect{k},\vect{k'}),
\end{equation}
with $\vect{k'}=\vect{k}+\vect{q}$. Notice that in contrast with Eq. (\ref{Eq.Polarizability}), when considering a single valley the energy dispersions $E_{k\alpha}=\alpha v_0 k$ only have one index $\alpha$ and the scattering probability $F_{\alpha\alpha'}(\vect{k},\vect{k'})=|\langle\Psi_{k'\alpha'}|\Psi_{k,\alpha}\rangle|^2$ is calculated from the single-valley eigenvectors $|\Psi_{k\alpha}\rangle=\frac{1}{\sqrt{2}}(1,\alpha e^{-i\theta_k})^T$, with $\theta_k=\tan^{-1} (k_y/k_x)$. One obtains, $F_{\alpha\alpha'}(\vect{k},\vect{k'})=\frac{1}{2}[1+\alpha\alpha'\cos(\theta_k-\theta_{k'})]$ and in order to leave the expression in terms of $q$ we use $\cos(\theta_k-\theta_{k'})=(k+q\cos\varphi)/|\vect{k}+\vect{q}|$, with $\varphi=\theta_q-\theta_k$, leading to
\begin{equation}\label{Eq.App.F_single_valley}
    F_{\alpha\alpha'}(\vect{k},\vect{q})=\frac{1}{2}\bigg(1+\alpha\alpha'\frac{k+q\cos\varphi}{|\vect{k}+\vect{q}|}\bigg).
\end{equation}
This is the single-valley scattering probability.

The single-valley polarizability $\Pi_{v_0}^g(\omega,q)$ in Eq. (\ref{Eq.App.P_single_valley}) has a well-known analytical solution, but the expression is quite complicated \cite{Sarma2007,Wunsch,Peres}. The calculation of the single-valley polarizability for massive (rather than massless) Dirac Fermions, $\Pi_{m_0}^g(\omega,q)$, is completely analogous and also has a well-known solution \cite{Pyatkovskiy2008}.

The eigenvectors of $H_K$ for the Kek-Y phase are $|\Psi_{k\alpha}^\beta\rangle=\frac{1}{2}(\beta,\alpha\beta e^{i\theta_k},\alpha e^{-i\theta_k},1)^T$ \cite{Herrera2020}. The scattering probability $F_{\alpha\alpha'}^{\beta\beta'}(\vect{k},\vect{q})=|\langle\Psi_{k'\alpha'}^{\beta'}|\Psi_{k\alpha}^\beta\rangle|^2$, with $\vect{k'}=\vect{k}+\vect{q}$ is thus given by
\begin{equation}
  F_{\alpha\alpha'}^{\beta\beta'}(\vect{k},\vect{q})=\frac{1}{4}[1+\alpha\alpha'\cos(\theta_k-\theta_{k'})][1+\alpha\alpha'\beta\beta'\cos(\theta_k-\theta_{k'})].
\end{equation}
Using again $\cos(\theta_k-\theta_{k'})=(k+q\cos\varphi)/|\vect{k}+\vect{q}|$ leads to 
\begin{equation}\label{Eq.App.F_plus}
   F_{\alpha\alpha'}^{+}(\vect{k},\vect{q})=\frac{1}{2}\bigg(1+\alpha\alpha'\frac{k+q\cos\varphi}{|\vect{k}+\vect{q}|}\bigg)-\bigg(\frac{q\sin\varphi}{2|\vect{k}+\vect{q}|}\bigg)^2,
\end{equation}
\begin{equation}\label{Eq.App.F_m}
   F_{\alpha\alpha'}^{-}(\vect{k},\vect{q})=\bigg(\frac{q\sin\varphi}{2|\vect{k}+\vect{q}|}\bigg)^2.
\end{equation}
We identify the first term on the right side of Eq. (\ref{Eq.App.F_plus}) as the single-valley polarizability of Eq. (\ref{Eq.App.F_single_valley}). We can therefore resume Eqs. (\ref{Eq.App.F_plus}) and (\ref{Eq.App.F_m}) as in Eq. (\ref{Eq:Fq_H1}).

Substituting Eq. (\ref{Eq:Fq_H1}) into Eq. (\ref{Eq.Polarizability}) and summing over the $\beta,\beta'$ indices allows to separate the polarizability of the Kek-Y phase as
\begin{multline}
    \Pi_Y(\omega,q)=-g_s\sum_{\alpha,\alpha'}\int \frac{d^2k}{4\pi^2}\frac{f_{k\alpha}^\beta-f_{k'\alpha'}^{\beta'}}{E_{k\alpha}^+-E_{k'\alpha'}^{+}+\omega^+}F_{\alpha,\alpha'}(\vect{k},\vect{q})\\-g_s\sum_{\alpha,\alpha}\int \frac{d^2k}{4\pi^2}\frac{f_{k\alpha}^\beta-f_{k'\alpha'}^{\beta'}}{E_{k\alpha}^--E_{k'\alpha'}^{-}+\omega^+}F_{\alpha,\alpha'}(\vect{k},\vect{q})\\+g_s\sum_{\alpha,\alpha'\beta\beta'}\int \frac{d^2k}{4\pi^2}\frac{f_{k\alpha}^\beta-f_{k'\alpha'}^{\beta'}}{E_{k\alpha}^\beta-E_{k'\alpha'}^{\beta'}+\omega^+}\bigg(\frac{q\sin\varphi}{|\vect{k}+\vect{q}|}\bigg),
\end{multline}
with $E_{k\alpha}^\beta=\alpha v_\beta k$ [given by Eq. (\ref{Eq.E_kek})].
The first two terms are identified with the single-valley polarizability of Eq. (\ref{Eq.App.P_single_valley}) for velocities $v_\pm=v_0\pm\Delta v_0$ and expressed as $\Pi_{v_\pm}^g(\omega,q)$ in Eq. (\ref{Eq.Polarizability_kek}) while the last term, which produces the signature at $\omega_M$, is expressed as $\Pi_{v_M}^M(\omega,q)$.
\\~\\
\section{Optical conductivity and activation frequencies}\label{App.Sigma}
The optical conductivity $\tilde{\sigma}(\omega)$ in a single valley can be obtained from the polarizability as \cite{Peres},
\begin{equation}\label{Eq.App.Sigma}
    \tilde{\sigma}(\omega)=\lim_{q\rightarrow0}i\frac{-\pi\omega}{2q^2}\Pi_{v_0}^g(\omega,q).
\end{equation}
For the Kek-Y phase, the signatures at $\omega_\pm$ in the optical conductivity (shown in Fig. \ref{Fig:Sigma_abc}a) can be traced to the 
$\Pi_{v_\pm}^g(\omega,q)$ terms in the polarizability, and thus identified as the activation frequencies of each specie of quasiparticle. A simple way to see this is by considering first that, in pristine graphene, the activation frequency for the Dirac fermions with Fermi velocity $v_0$ is $\omega=2\mu$, and this leads the optical conductivity to be given by a step function $\tilde{\sigma}(\omega)\sim\Theta(\omega-2\mu)$ \cite{Peres}. On the other hand, in the Kek-Y phase [see Eq. (\ref{Eq.Polarizability_kek})] the first two terms, $\Pi_{v_\pm}^g$, are given by the same single-valley polarizability of Eq. (\ref{Eq.App.P_single_valley}), only with a shift in the Fermi velocity $v_0\rightarrow v_\pm=v_0(1\pm\Delta)$. Note that $\mu= v_0 k_F$, and therefore scaling $v_0\rightarrow v_0(1\pm\Delta)$ also scales $\mu$ as $\mu\rightarrow\mu(1\pm\Delta)$. This then shifts the activation frequency as $\omega\rightarrow 2\mu(1\pm\Delta)$, which indeed coincides with the activation frequencies $\omega_\pm$ in Fig. \ref{Fig:Sigma_abc}a. An analogous analysis can be done for the signatures in Fig. \ref{Fig:Sigma_abc}b corresponding to the Kek-M phase.

\bibliography{Refs_Beating}
\end{document}